\documentstyle[psfig]{mn}
\begin{document}

\title[Shell formation in NGC 474 and NGC 7600]{Imaging of the Shell Galaxies 
NGC 474 and NGC 7600, and Implications for their Formation}

\author[A. J. Turnbull, T. J. Bridges and D. Carter]{A. J. Turnbull,$^1$
\thanks{Email: ajt@star.herts.ac.uk} T. J. Bridges$^2$\thanks{Email: 
tjb@aaoepp.aao.gov.au} D. Carter$^3$\thanks{Email: dxc@astro.livjm.ac.uk}\\
$^1$Department of Physical Sciences, University of Hertfordshire, Hatfield, 
College Lane, AL10 9AB, UK.\\
$^2$Institute of Astronomy, Madingley Road, Cambridge, CB3 OE3, UK.\\
Current address: Anglo-Australian Observatory, PO Box 296, Epping, 
NSW 1710, Australia.\\
$^3$Liverpool John Moores University, Astrophysics Research Institute, Twelve 
Quays House, Egerton Wharf, Birkenhead, L41 1LD\\}
\maketitle
\begin{abstract}
We present photometric observations of two shell galaxies, 
NGC 474 and NGC 7600. We examine the photometric 
colours and surface brightnesses of the shells and their host galaxies,
and the isophotal parameters of each galaxy. 
In the case of NGC 474, we find that the shell formation is consistent with a 
merger origin although it is possible that the close companion NGC 470 
is contributing to the shell system via mass transfer.
NGC 7600 exhibits shell geometry and colours which also favour 
a merger origin.
\end{abstract}


\section{Introduction}

In recent years it has become clear that elliptical galaxies are host 
to much fine structure. Perhaps the most dramatic fine structures 
observed are the edge-brightened arcs of stars surrounding 
and existing within the envelopes of some ellipticals. Originally classified 
as peculiar galaxies by Arp (1966), but later rediscovered and renamed 
`shell galaxies' by Malin \& Carter (1980), the exact nature and origin of 
these features still remains unclear. 

The publication of Malin and Carter's (1983) catalogue of 137 shell galaxies, 
found in the SRC/ESO Southern Sky Survey by eye, showed 
that about {17\%} of field ellipticals were surrounded by shells. 
Seitzer \& Schweizer (1990) examined 74 elliptical and SO galaxies,
north of $\delta = -15\deg$, brighter than $B_{T} = 13.5$, and 
nearby $(cz > 4000 {km\,s^{-1}})$.  They found shells (referred to by 
Schweizer as ripples, but the name is interchangeable) in more than 
half of the field ellipticals and in one third of the field SOs. All found 
the frequency of shells in spirals to be almost negligible. Clearly shells 
are common in elliptical and S0 galaxies.
Shell formation and evolution needs to be understood 
if a complete picture of the formation and evolution of the galaxies they 
inhabit is to be obtained.

A variety of models have been proposed to form shells (see 
reviews by Prieur 1990 and Carter 1998).
In the minor merger models (Quinn 1984; Dupraz \& Combes 1986; 
Hernquist \& Quinn 1988, 1989) shells are formed from the stars of the 
secondary galaxy, which merges within the primary.  The shells are a 
result of the phase wrapping
(low orbital angular momentum encounters) or spatial wrapping (high 
orbital angular momentum encounters) of stars belonging to the 
secondary. The nature of the secondary can also affect the 
resulting shell system.  Not surprisingly then, combining the small 
cross section for nearly radial encounters with the many types of 
accretion candidates leads to a wide variety of observed shell 
morphologies.  Heisler \& White (1990) pointed out the shortcomings of 
the three body approach, and 
used an accretion scenario to study the effect of tidal 
disruption on the secondary.  They found that although the positions of the 
shells can be reproduced using test particles, in order to accurately 
reproduce the population of the shells, a self-consistent treatment of the 
disruption is essential.  Dynamical friction against the primary galaxy
is another important ingredient,
and is believed to allow shells to be produced deep within the 
potential well.  Dupraz \& Combes (1987) investigated this analytically.
They proposed that the stars least tightly bound to the companion (probably
a large fraction of its mass) are pulled away by tidal forces during the 
first passage close to the main galaxy centre. Afterwards, the surviving 
companion is braked by friction as it orbits through the galaxy. More
stars are `peeled' by tidal forces launched with continuously lower and lower
energies. The innermost shells develop at the end, when merging is almost
complete.

Whether major disk-disk mergers can also form shells was asked by 
Schweizer (1980).  Hernquist \& Spergel (1992) reproduced shell-like structures
from such a scenario and Hibbard \& Mihos (1995) carried out 
major merger modelling of NGC 7252 and found that
infalling tidal material may be able to 
create shells; however the simulation was not evolved to this latter stage. 
The 
discovery by Balcells (1997) of two very faint, opposed tidal tails, 
usually considered as a signature of a disk-disk merger, in the shell 
elliptical NGC 3656 lends some weight to the theory. 

Thomson \& Wright (1990) put forward a weak interaction model (WIM) as 
an alternative to the above merger models.  The shells are 
produced from a one-armed 
spiral density wave induced within a dynamically cold 
`thick disk' component of the 
primary galaxy during a fly-by interaction (not a merger) with a 
secondary companion; the shell-forming stars are 
initially on circular but not co-planar orbits. Although hot systems such 
as ellipticals would not be expected to contain such a population of 
stars on circular
orbits, the effectiveness of the test particle model in reproducing 
so efficiently the shell distributions of NGC 3923 and 0422-476, 
particularly the inner shells (Thomson 1991), lends credence to the 
WIM.  However, Carter, Thomson \& Hau (1998) have recently reported 
minor axis rotation in NGC 3923, which implies that the underlying 
potential is in fact prolate. Together with the high velocity dispersion 
of the galaxy,
this argues against the existence of a thick disk as required by the WIM. 
Carter, Thomson \& Hau (1998) include a full discussion on these points. 

Photometry offers a simple yet effective way of constraining models of 
shell formation.
Assuming that
no significant star formation is induced by the one-armed spiral density
wave, a 
natural prediction of the WIM is that the shell colours should follow 
the colour gradient of the host galaxy.  Accretion of a cold, disk galaxy in 
its early stages of merging should show a marked colour 
difference between the red host elliptical and the bluer companion. As the 
disk ages the colours will redden and the colour difference will be less 
pronounced. Other secondary candidates are low mass elliptical galaxies, 
which should still show a slight difference in colour as low mass elliptical 
galaxies are bluer on average than larger E type 
galaxies (Visvantha \& Sandage 1977). 

Previous 
photometric studies of shell galaxies (Carter, Allen \& Malin 1982; Fort
 et al 1986; 
Schombert \& Wallin 1987; Forbes et al 1995) have in general shown that the
 shells 
are similar in colour or slightly bluer than the underlying galaxy,
although uncertainties in the shell colours and geometry are large.
The motivation of this study is to provide high signal to noise data 
capable of allowing clear distinctions to be drawn between 
formation models, and also allowing us 
to learn more about the stellar population(s) the shells contain. 

The surface brightness of the shells, when compared to the 
underlying galaxy, should provide a further test of the models. In the WIM the 
shells are a density wave excited in a component of the galaxy, and 
will be an approximately constant fraction of the surface brightness of 
this component. It is more difficult to predict the expected distribution 
in a merger scenario for reasons already mentioned. Through their 
investigation, Dupraz \& Combes (1987) suggested that the outer shells should
contain the highest surface brightness. 
It is difficult to know how to interpret galaxy isophotal data.
Bender et al (1988), Bender (1997) and references therein used the fourth
 cosine term
 (B$_4$) in the Fourier expansion of the azimuthal variation to quantify the 
isophotal deviations from pure ellipses. It is thought that pointy 
isophotes (positive B$_4$) are associated with discs (Nieto et al 1991) 
whereas galaxies with box-shaped isophotes (negative B$_4$) are thought 
to have been produced by a merger or interaction (Binney \& Petrou 1985, 
Bender 1990). Shells are expected to produce boxy isophotes as they 
follow isopotential not isodensity surfaces and isopotentials are always 
rounder (Dupraz \& Combes 1986). It is interesting to 
ask whether the correlations between boxiness and other evidence for 
interactions can be entirely explained by the presence of shells as suggested
by Forbes \& Thomson (1992). Changes in the 
semi-major axis position angle with radius can suggest triaxiality.  

We present $B-R$ colours and surface brightnesses for NGC 474 and NGC 7600,
and their 
shells.  We also present an isophotal analysis for each galaxy, as well as an 
investigation into associated fine structure and environment. 
Observations are discussed in \S2, data reduction in \S3 and 
results in \S4. We analyse our findings in the context of the 
shell formation models in \S5 and summarise our conclusions in \S6.

\section{Observations}

Broadband {\it B} images of NGC 474 were taken at the prime focus  
of the 4.2m William Herschel Telescope in December 1994.  We 
used a 1024 $\times$ 1024 Tektronix thinned CCD, giving
a field of 7.2 $\times$ 7.2 arcminutes with 0.42 arcsec/pixel. 
NGC 474 
fit comfortably onto the chip, although there is some slight clipping
of the outer shell, lying 190 arcseconds from the galaxy 
centre.  Broadband {\it R} 
and narrowband {\it H${\alpha}$} images of NGC 474 and {\it B, R} for NGC 7600 
were obtained at the 2.5m Isaac Newton Telescope in November 1996.  Again a 
Tektronix CCD 
was used at prime focus, giving a field of view of 11.5 $\times$
10.6 arcminutes and an image
scale of 0.59 arcsec/pixel.  
The seeing was 
$\sim$ 1.5 arcsec for each run and filter. The total integration times 
are listed in Table 1. Although included in the table, the {\it H${\alpha}$} 
image is not presented in this paper. 
When a scaled 
$R$-band image was subtracted from the {\it H${\alpha}$} image,  
no traces of {\it H${\alpha}$} emission were visible.

\begin{table*}
\caption{Summary of the Observations.   Positions,
recession velocities, and B$_T$ magnitudes
are taken from the NASA/IPAC Extragalactic Database (NED).}
\begin{tabular}{cccccccc} \hline
Galaxy & Filter & RA & Dec & $V_{r}$ & $B_{T}$  & 
Integration Time & Telescope \\  
 & & (1950)& (1950)  & (km/sec) & & (sec) & \\ \hline
NGC 474 & {\it R} & 01$^h$17$^m$31.7$^s$ & +3$^{\circ}$9$\arcmin$17$\arcsec$ &
 2372  & 
12.37 & 2700 & INT \\
NGC 474 & {\it H${\alpha}$} & & & & & 9000& INT \\
NGC 474 & {\it B} & & & & & 3600 &  WHT	\\
NGC 7600 & {\it R} & 23$^h$16$^m$18.2$^s$ & -7$^{\circ}$51$\arcmin$11$\arcsec$
 & 3436 
&12.91& 
2700 &  INT   \\
NGC 7600 & {\it B}& &  & & &5400 &  INT\\ \hline
\end{tabular}
\end{table*}

\section{Data Reduction}

Standard {\sc iraf} data reduction techniques were used. After debiasing, 
our images were divided by a normalised flat field in the appropiate 
pass band, each obtained from a number of exposures taken during twilight. 
Each passband's 
individual images, aligned to within a fraction of a pixel using 
{\sc imalign}, 
were then combined using {\sc imcombine}. 
This employed a median algorithm, scaled to the mode and was
very efficient at removing the cosmic rays.  The seeing was $\sim$ 1.5
arcsec for 
all images, and this as well as the accuracy of 
the aligning was checked by examining the FWHM of several stars in the 
frame before and after combining. When aligning the combined images with 
each other the same procedure was used as before except in the case of 
NGC 474, where as mentioned earlier the {\it B} and {\it R} images were taken 
on different telescopes. It was therefore necessary to correct for different 
chip scales and pixel sizes. The packages {\sc geomap} and {\sc geotran} 
were used for this. {\sc geomap} computes the transformation required to map
 the 
coordinate system of the reference image to that of the 
input image,
using over a dozen stars common to 
each frame, while
{\sc geotran} performs the transformation. Total flux was 
conserved and the accuracy checked as before. 

\subsection{Photometric Calibration}

Calibration was performed using the {\sc phot} package in {\sc iraf}. The 
zero point for each night was found using a number of 
Landolt (1992) standard stars 
taken throughout the night; it was found that colour terms were not required.
Mean extinction values provided by the Carlsberg Meridian Telescope 
were used to correct for atmospheric extinction,  
based on many observations of 
standard stars throughout the night. Although small, galactic extinction 
has been corrected for using the interstellar extinction law given 
by Rieke \& Lebofsky (1985).  The zero point was calculated for an 
individual galaxy frame and then recalibrated to the combined 
galaxy frame. The resulting zero points of the combined galaxy frames 
are displayed in Table 2. The difference in {\it R} zero points 
between NGC 474 and NGC 7600
is due to the different scaling used to produce the final composite galaxy
images. 

\begin{table}
\caption{Calibrated zero points.}
\begin{tabular}{ccc} \hline
Galaxy and Filter & Zero Point & rms $(\frac{\sigma}{\sqrt{N}})$ \\ \hline 
        NGC 474 {\it R} & 25.081 & 0.02 \\
	NGC 474 {\it B} & 25.768 & 0.01 \\
	NGC 7600 {\it R} & 24.895 & 0.01 \\
	NGC 7600 {\it B} & 25.175 & 0.01  \\ \hline   
\end{tabular}
\end{table}

\begin{figure}
\psfig{{figure=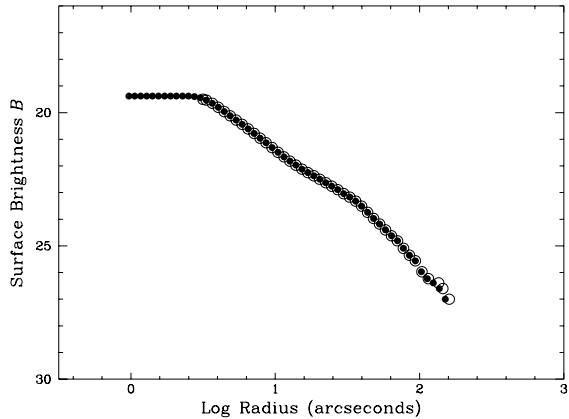},width=8cm,angle=270}
\caption{Comparison of NGC 474 {\it B} photometric calibration with 
Schombert \& Wallin (1987). Our data are overplotted as filled circles.}
\end{figure}

\begin{figure}
\psfig{{figure=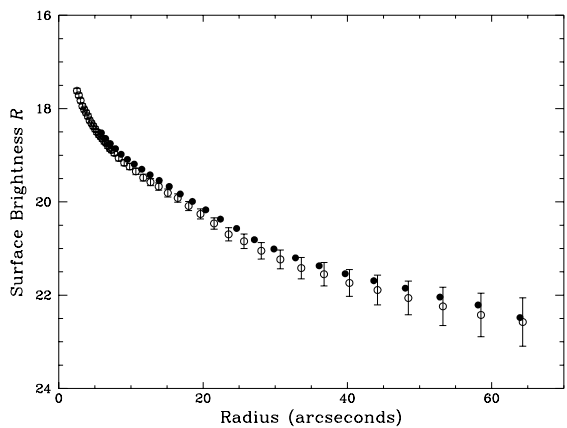},width=8cm,angle=270}
\caption{Comparison of NGC 7600 {\it R} photometric calibration with 
Penereiro et al., (1994) converted from Gunn {\it r} to Kron-Cousins {\it R}.
 Our data are overplotted as filled circles.}
\end{figure}

To test the reliability of our calibration, we have compared the surface 
brightness profiles of our galaxies with those published by other authors. 
Shown in Figure 1 is the $B$ surface brightness profile for NGC 474 taken from 
Schombert \& Wallin (1987). Our results, overplotted as filled circles, are in 
very good agreement.  For NGC 7600 we have compared our data with that of 
Penereiro et al., (1994).  Shown in Figure 2 as open circles with error 
bars is the surface brightness of NGC 7600 transformed from the Gunn {\it r}
 filter 
as 
deduced by Penereiro et al into our Kron-Cousins {\it R}; our data are 
overplotted as filled circles, the 
symbol size being of the order of our error. We used the transformation 
given by Jorgensen (1994) to convert from the Gunn system to our Kron-Cousins 
photometric system.  There is a zeropoint difference of
$\sim$ 0.2 mag between our photometry and that of Penereiro et al., but
the agreement is satisfactory.

\subsection{Modelling the Underlying Galaxy}

Apart from rare cases such as the outer shells of NGC 474, shells are 
faint structures and difficult to detect since they are superimposed 
upon the bright background of the galaxy. The galaxy signal also contaminates 
the signal from a shell.  It is necessary therefore to 
produce and subtract off a model of the underlying galaxy.  First, the 
contribution of the sky background was estimated from the mean value of
several 10 $\times$ 10 pixel boxes placed in regions well away from the 
galaxy.  Our isophotal analysis made use of the 
{\sc ellipse} task within {\sc stsdas}\footnote{The Space Telescope Science 
Data Analysis System {\sc stsdas} is distributed by the Space Telescope 
Science Institute.}, which is based on the method described in 
detail by Jedrzejewski (1987).  We used a similar technique of 
ellipse fitting as that described by Forbes \& Thomson (1992). 
In the case of NGC 474, where the bright shells significantly affect the 
galaxy isophotes, a first model was run with the centre, position 
angle and ellipticity allowed to vary.  A second model was then run with 
the central position fixed at the value found from 
the first run.  The brightest 10\% of pixels in each ellipse 
were excluded, to avoid the brightest points in the shells. NGC 7600
was found to produce a residual image of comparable quality whether or not 
the parameters were fixed, and for this galaxy all parameters
were allowed to vary and no clipping was used. 

\begin{figure*}
\begin{center}
{\psfig{figure=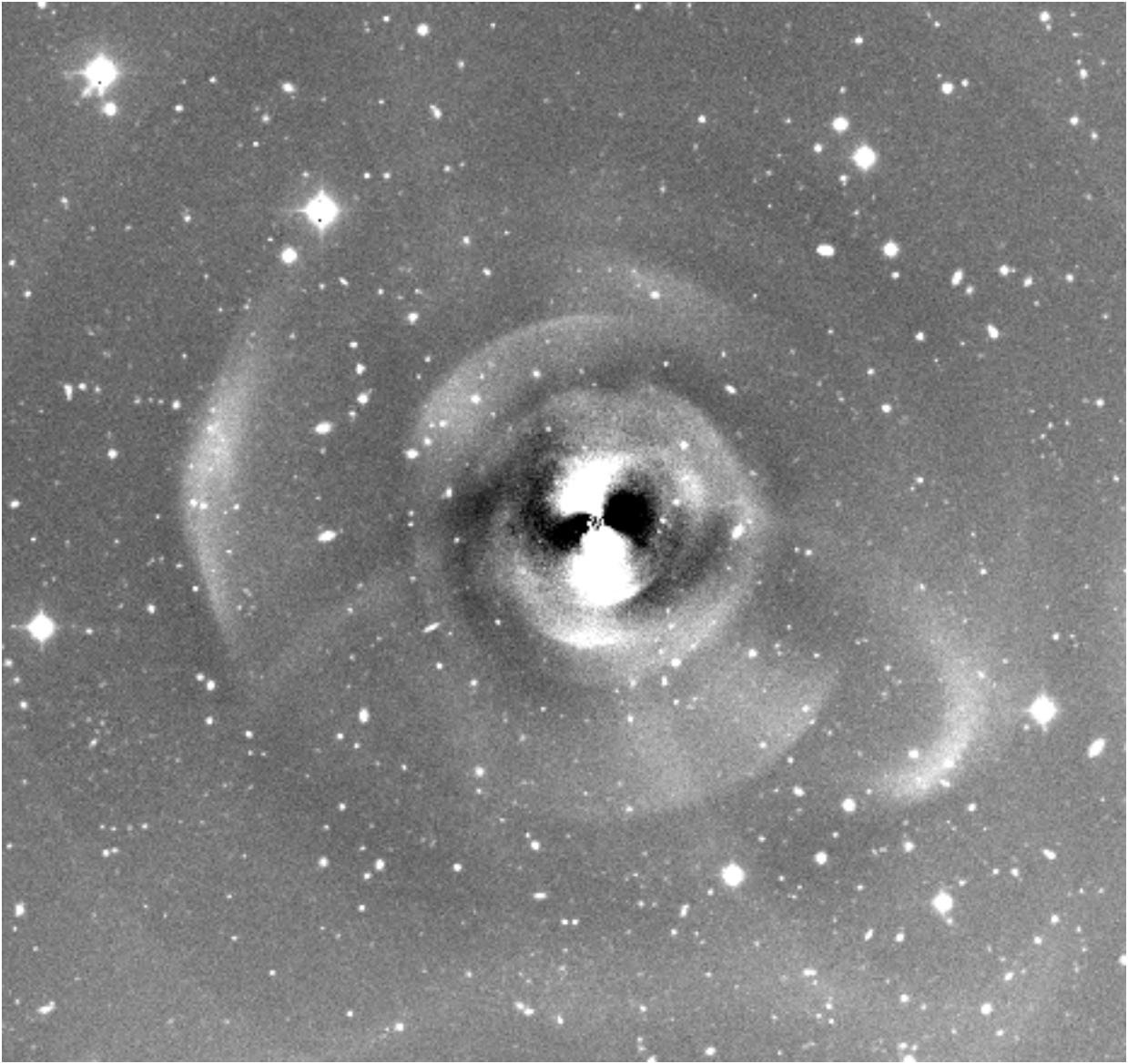}}
\caption{{\it R} band galaxy subtracted image of NGC 474. North is up and 
East is to the left. The eastern-most shell is 202 arcsec from 
the galaxy centre.  NGC 470 is located just off the frame, $\sim$ 300
arcsec West. The field of view is 9$\arcmin$.}
\end{center}
\end{figure*}

\begin{figure*}
\begin{center}
{\psfig{file=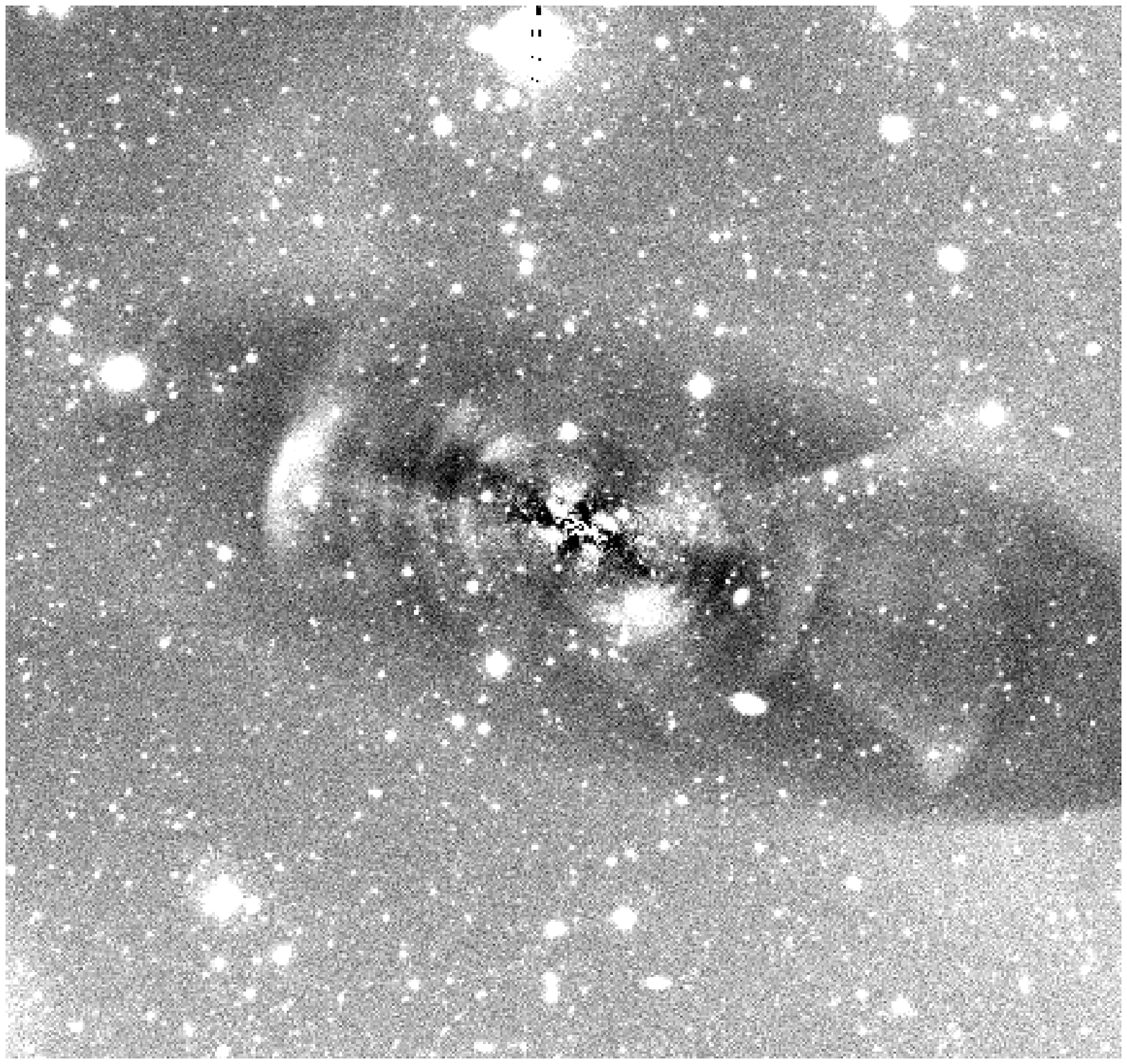}}
\caption{{\it R} band galaxy subtracted image of NGC 7600.  North is up and 
East is to the left. The dark oval shape is an artifact of the 
subtraction process. The eastern-most shell lies 215 arcsec away from 
the galaxy centre. The field of view is 9$\arcmin$.}
\end{center}
\end{figure*}

\subsection{Aperture Photometry of Shells}

First we identified the regions of the residual map where shells are present. 
A 10 $\times$ 10 pixel box was placed on the shell, being 
careful to avoid foreground stars, globular clusters and any other 
unwanted features.  On either side of shell the task was repeated in 
order to deduce the local sky+galaxy background.
The procedure was repeated along the shell, the number of times depending
upon the size of shells. Typically this ranged from 5 to 15 independant
boxes.
The 
corresponding shell surface brightness $(SB)$ was thus deduced for 
each aperture 
using the formula:

\begin{equation}
	SB = -2.5*log(\frac{counts}{itime}) + m0
\end{equation}

where $counts$ are per square arcsecond, $itime$ is the exposure time in 
seconds, and $m0$ the zero point for that filter. This was repeated for the 
other filter using identical positions, and the colour indices calculated. 
After checking for possible colour gradients, the mean value for the 
shell was determined.  Any remaining spurious features were re-examined, 
and those apertures lying further than two $\sigma$ away from the 
mean were rejected and the mean recalculated.  The final shell surface 
brightnesses were calculated by averaging the surface brightnesses 
calculated for each aperture along the shell.

\subsubsection{Error Analysis}

The low surface brightness of the shells and the high signal from the 
galaxy and sky lead to some uncertainty in the final shell 
magnitude due to random measurement errors.  We adopted the following 
procedure to calculate this uncertainty.
The value $\sigma_{D}$ was calculated for one of the band passes:

\begin{equation}
	\sigma_{D}=\sqrt{\sigma_{shell}^{2}+\frac{\sigma_{sky1+galaxy1}^{2}+
\sigma_{sky2+galaxy2}^{2}}{2}}
\end{equation}

These are the associated standard deviations of the values returned from the 
10 $\times$ 10 pixel box placed on and to each side of the shell. The procedure
was repeated for all apertures along the shell in that filter and the 
standard error for the whole shell calculated using (assuming we are in the $R$
Band):

\begin{equation}
	\sigma_{R}=\sqrt{\sum_{i=1}^{N}\frac{\left( \frac{\sigma_{Di}}
{\sqrt{n}} \right)^{2}}{N}}
\end{equation}

Here, $N$ is the number of apertures along the shell, which was typically no 
less than five and reached fourteen for some of NGC 474's larger shells, and $n$ 
is the number of pixels in the aperture which is 100 in our case. The above process 
was repeated for the B filter and the total error calculated as:

\begin{equation}
	\sigma_{B-R}=\sqrt{\left( \frac{\sigma_{B}}{\overline{n}_{B}} \right)^{2}+
\left( \frac{\sigma_{R}}{\overline{n}_{R}} \right)^{2}+(0.01)^{2}+(0.02)^{2}}
\end{equation}  

$\overline{n}_{R}$ and $\overline{n}_{B}$ are the average counts for the 
shell in $R$ and $B$ respectively.  We have adopted zero-point errors 
of 0.01 and 0.02 for B and R respectively, for NGC 474 (see Table 2).
For the NGC 474 shell at 202 
arcseconds radius, we calculated $\sigma_{B-R}$ = 0.05. We further calculated 
that this shell had an average $B-R$ of 1.00 leading to a final value
of 1.00 $\pm{0.05}$.
We believe $\pm$ 0.05 is a realistic error when obtaining shell colours based 
on the techniques described in this paper. 

\section{Results}

In Tables 3 and 4 we list, for NGC 474 and NGC 7600 respectively,
the average radius, surface brightness, and $B-R$ colour of each shell.
We also present plots of $B-R$ and surface brightness for the shells
and host galaxies in this section.

\begin{table}
\caption{NGC 474 shell parameters. For details see text.}
\begin{tabular}{ccc} \hline
	 Shell Radius & Brightness  & $B-R$ \\ 
	 (arcseconds) & ( R mag/arcsec$^{2}$) & \\ \hline
	202 &	24.67	& 1.00 ${\pm 0.05}$\\ 
	192 &       24.98   & 0.91 ${\pm 0.05}$\\
	140 &       25.46   & 1.21 ${\pm 0.08}$\\
	121 &       25.61   & 1.10 ${\pm 0.09}$\\
	100 &       24.92   & 1.24 ${\pm 0.06}$\\
	74  &       25.29   & 1.34 ${\pm 0.08}$\\
	72  &       24.95   & 1.33 ${\pm 0.06}$\\
	63  &	25.45   & 1.16 ${\pm 0.08}$\\
	55  &	24.53   & 1.21 ${\pm 0.05}$\\
	50  &	24.91	& 1.44 ${\pm 0.08}$\\ \hline
\end{tabular}
\end{table}

\begin{table}
\caption{NGC 7600 shell parameters. Blank spaces indicate that although a 
shell was present, it was either dominated by spurious objects or the 
signal was too low for accurate photometry. For further 
details see text.}
\begin{tabular}{cccc} \hline
	Shell Radius & Brightness  & $B-R$ \\ 
	(arcseconds) & ( R mag/arcsec$^{2}$) & \\ \hline
	215 &	25.76	& 1.35 ${\pm 0.12}$\\ 
	151 &       24.91   & 1.35 ${\pm 0.06}$\\
	118 &       25.91     & 1.37 ${\pm 0.15}$\\
	112 &       26.36        & 1.52 ${\pm 0.24}$\\
	  97   &             &     		   \\
	92  &       26.29      & 1.60 ${\pm 0.21}$\\
	 79  &	              &  		   \\
	76  &       25.91            & 1.53 ${\pm 0.18}$\\
	67	& & \\
	58  &       25.05	        & 1.49 ${\pm 0.04}$\\
	56 & & \\	
	53  &	25.66        & 1.46 ${\pm 0.11}$\\
	49  &       25.33 	& 1.68 ${\pm 0.13}$\\
	46 & & \\
	41 & & \\
	 38    &            &                  \\ \hline
\end{tabular}
\end{table}

\begin{figure}
\psfig{{figure=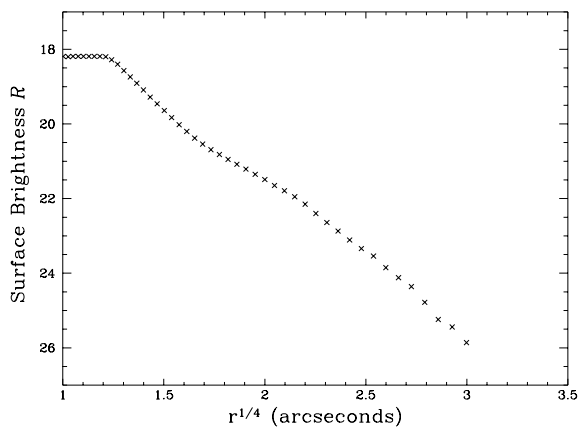},width=8cm,angle=270}
\psfig{{file=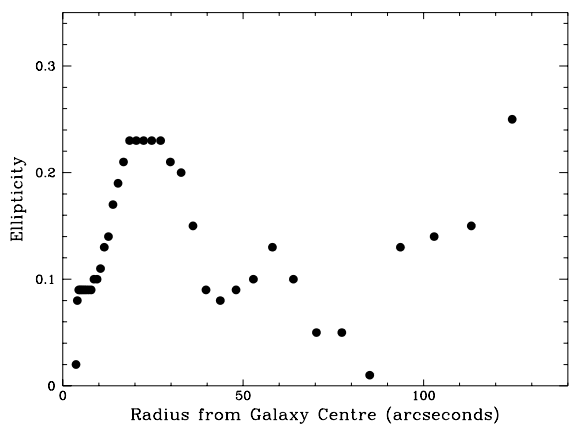},width=8cm,angle=270}
\caption{({\it 5a top}) NGC 474 $R$ surface brightness versus $r^{1/4}$ and
({\it 5b bottom}) NGC 474 ellipticity versus radius. The errors are of the 
order of the symbol size.}
\end{figure}

\begin{figure}
\psfig{{figure=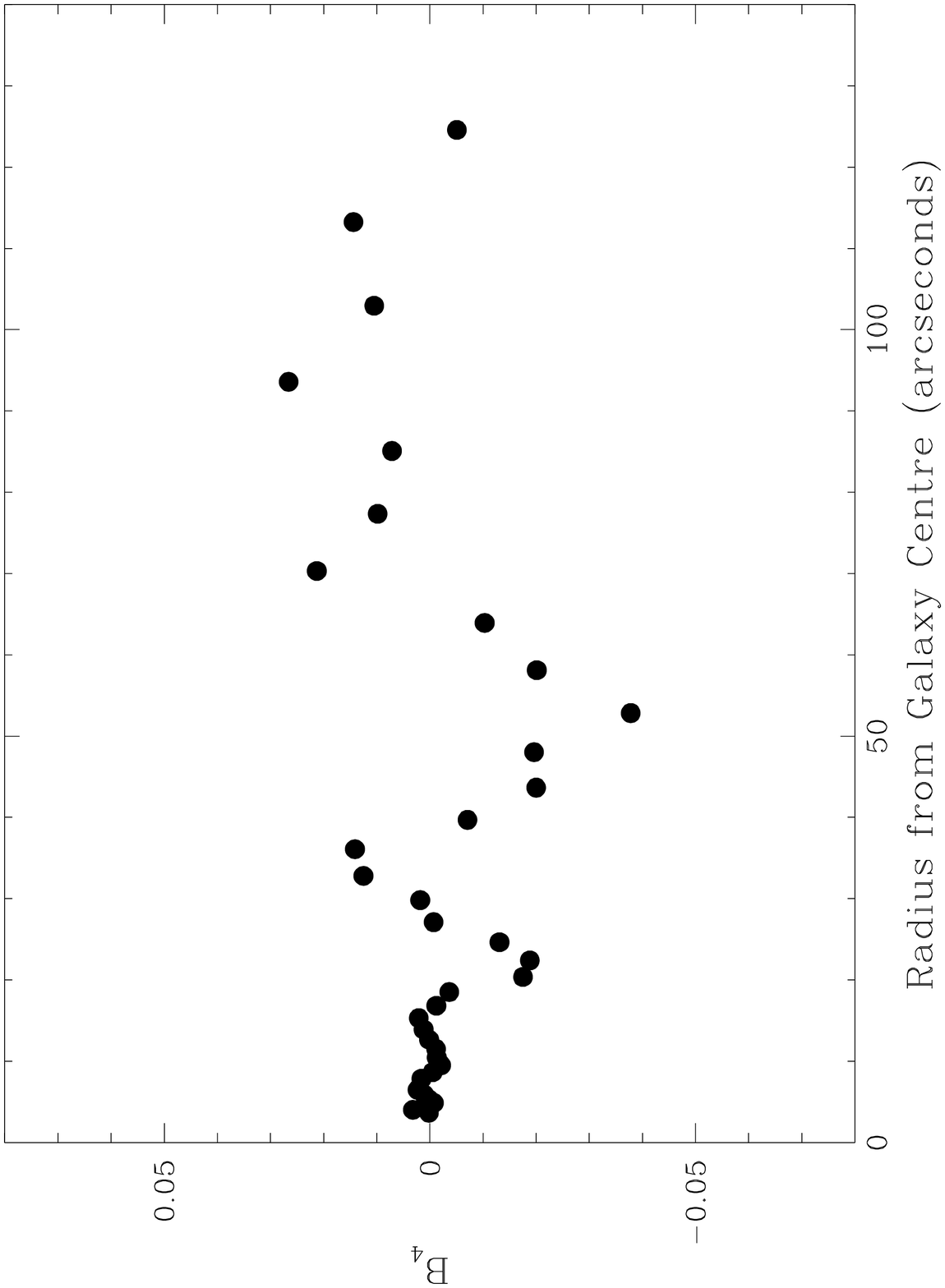},width=8cm,angle=270}
\psfig{{figure=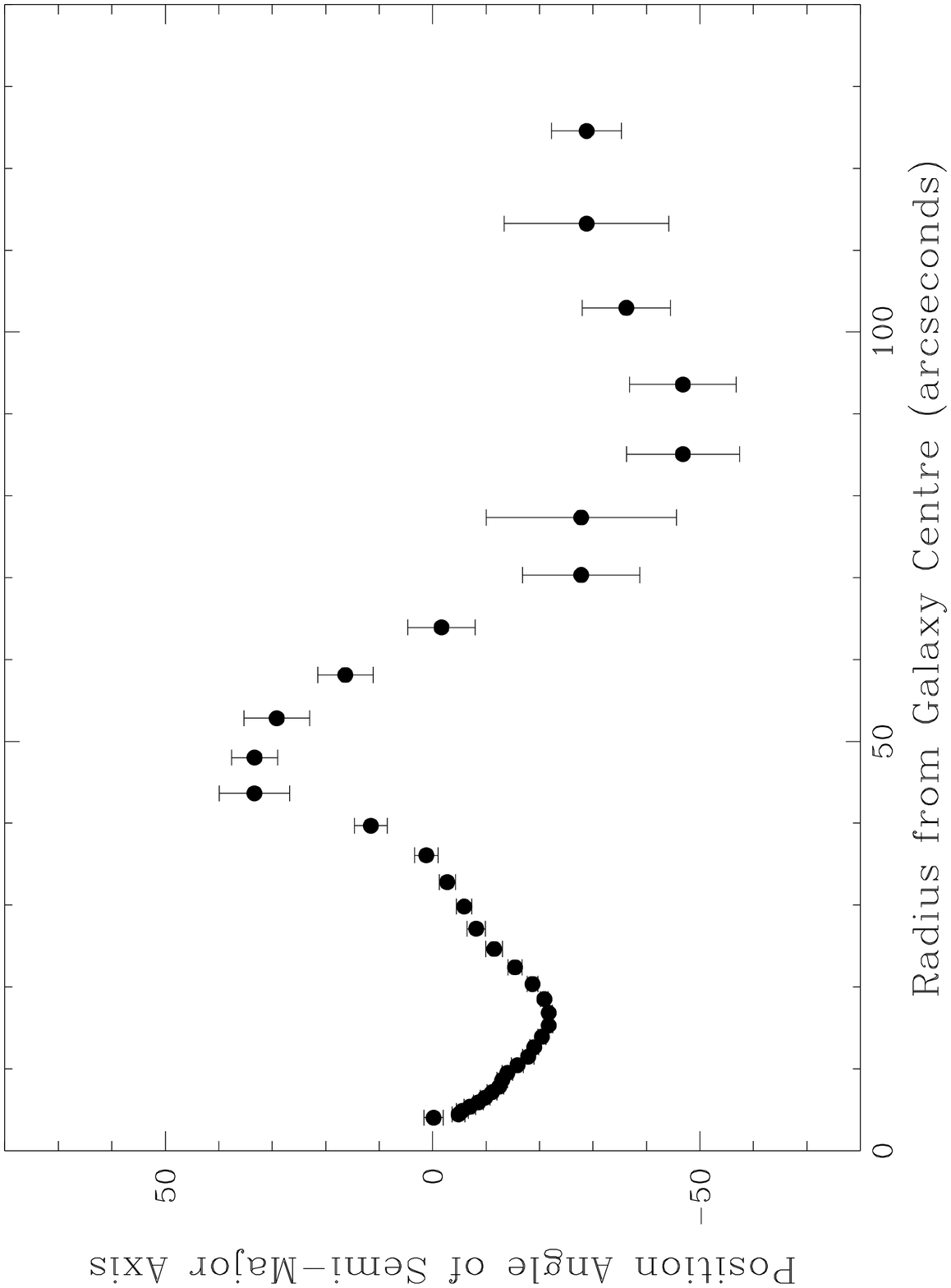},width=8cm,angle=270}
\caption{({\it 6a top}) NGC 474 cos$B_{4}$ term versus radius and ({\it 6b 
bottom}) 
NGC 474 position angle of the semi-major axis versus radius. The errors are of
the order of the symbol size in the absence of error bars.}
\end{figure}

\begin{figure}
\psfig{{figure=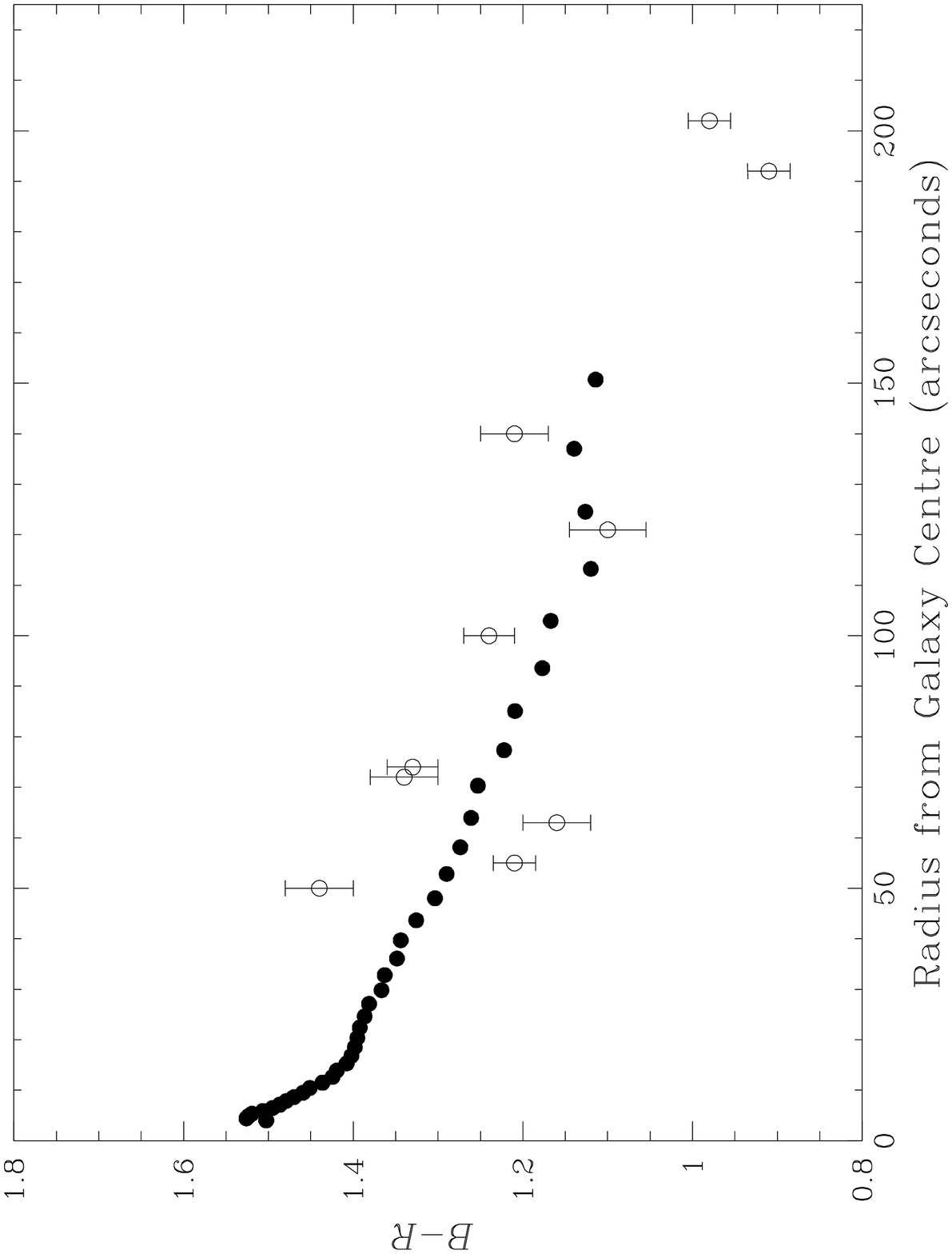},width=8cm,angle=270}
\psfig{{figure=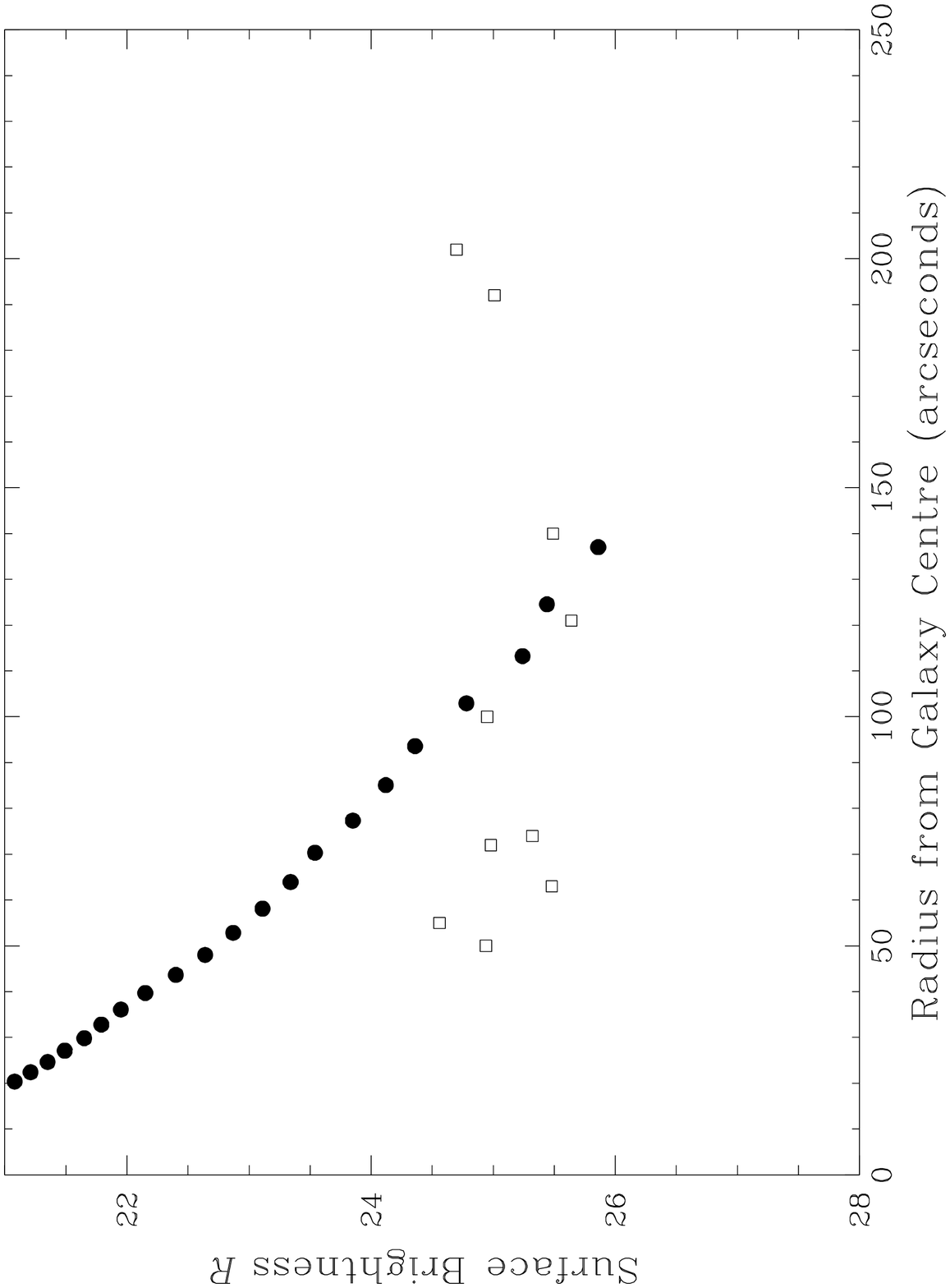},width=8cm,angle=270}
\caption{({\it 7a top}) NGC 474 $B-R$ of the galaxy and shells versus radius
 and ({\it 7b bottom}) the surface brightness of NGC 474 and shells versus 
radius. In all cases the galaxy data are represented as filled circles, the 
error being of the order of the symbol size. The shell data are open circles
with error bars.}
\end{figure}

\begin{figure}
\psfig{{figure=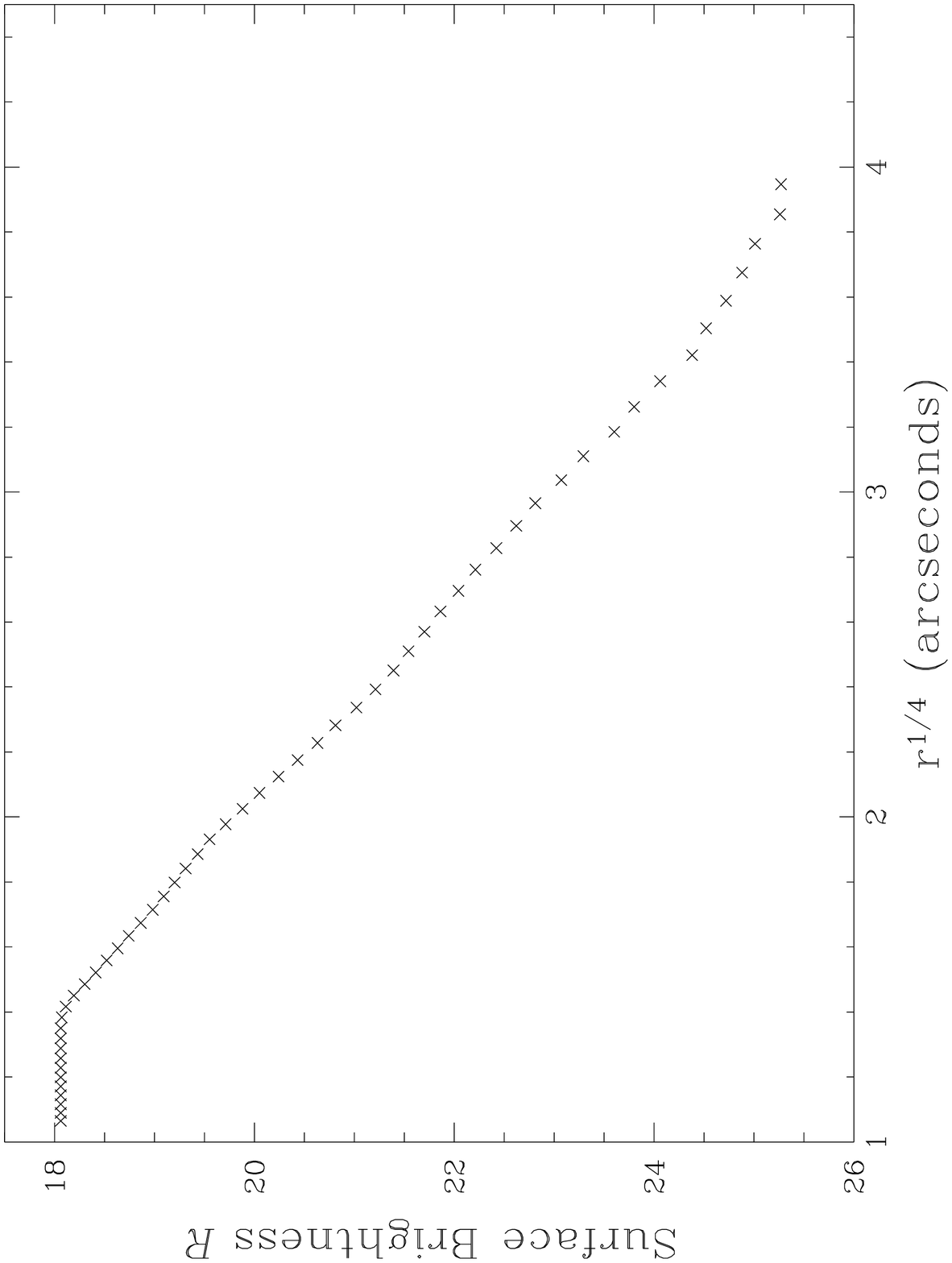},width=8cm,angle=270}
\psfig{{figure=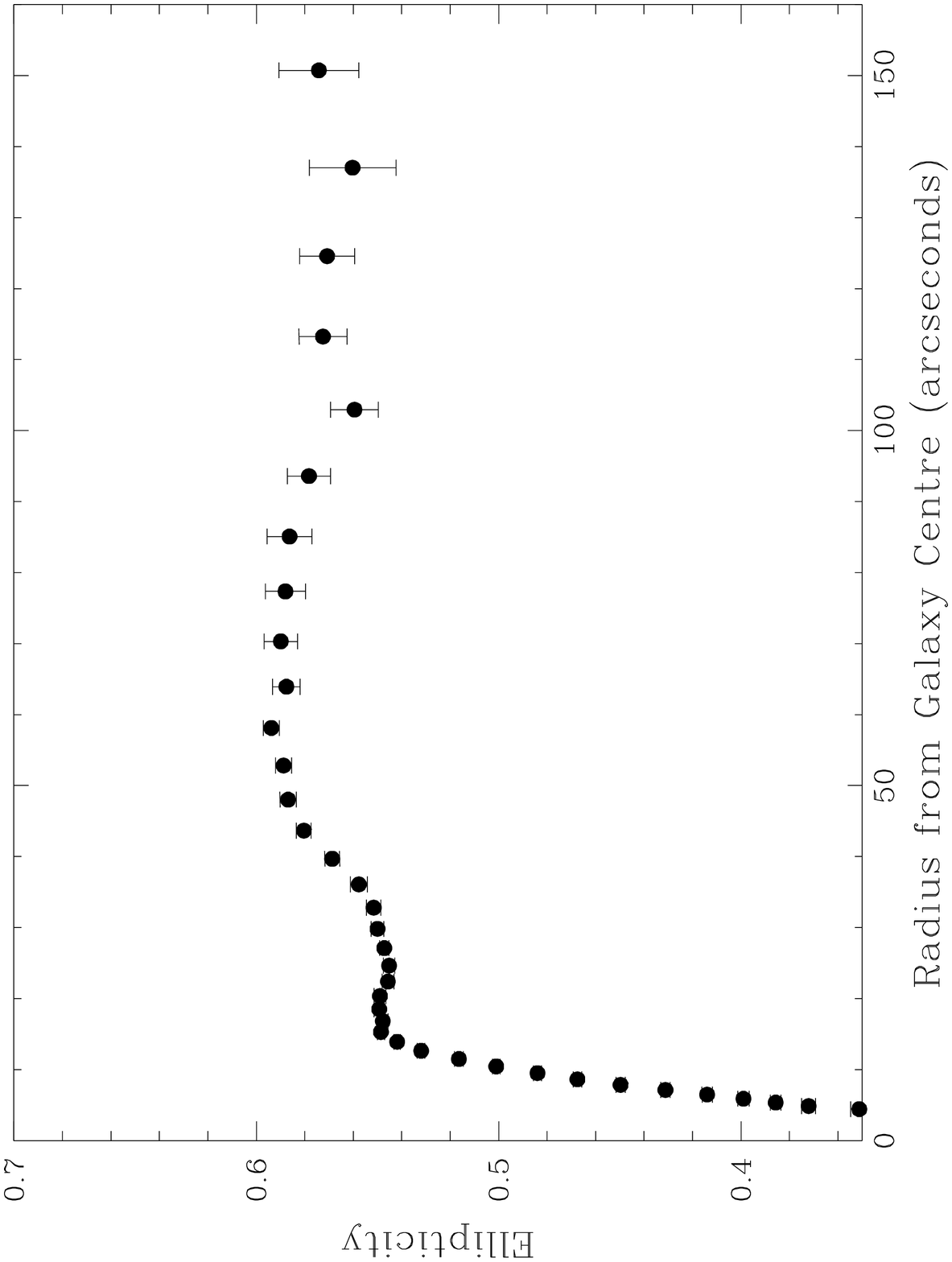},width=8cm,angle=270}
\caption{({\it 8a top}) NGC 7600 $R$ surface brightness versus $r^{1/4}$ 
and ({\it 8b bottom}) NGC 7600 ellipticity versus radius.}
\end{figure}

\begin{figure}
\psfig{{figure=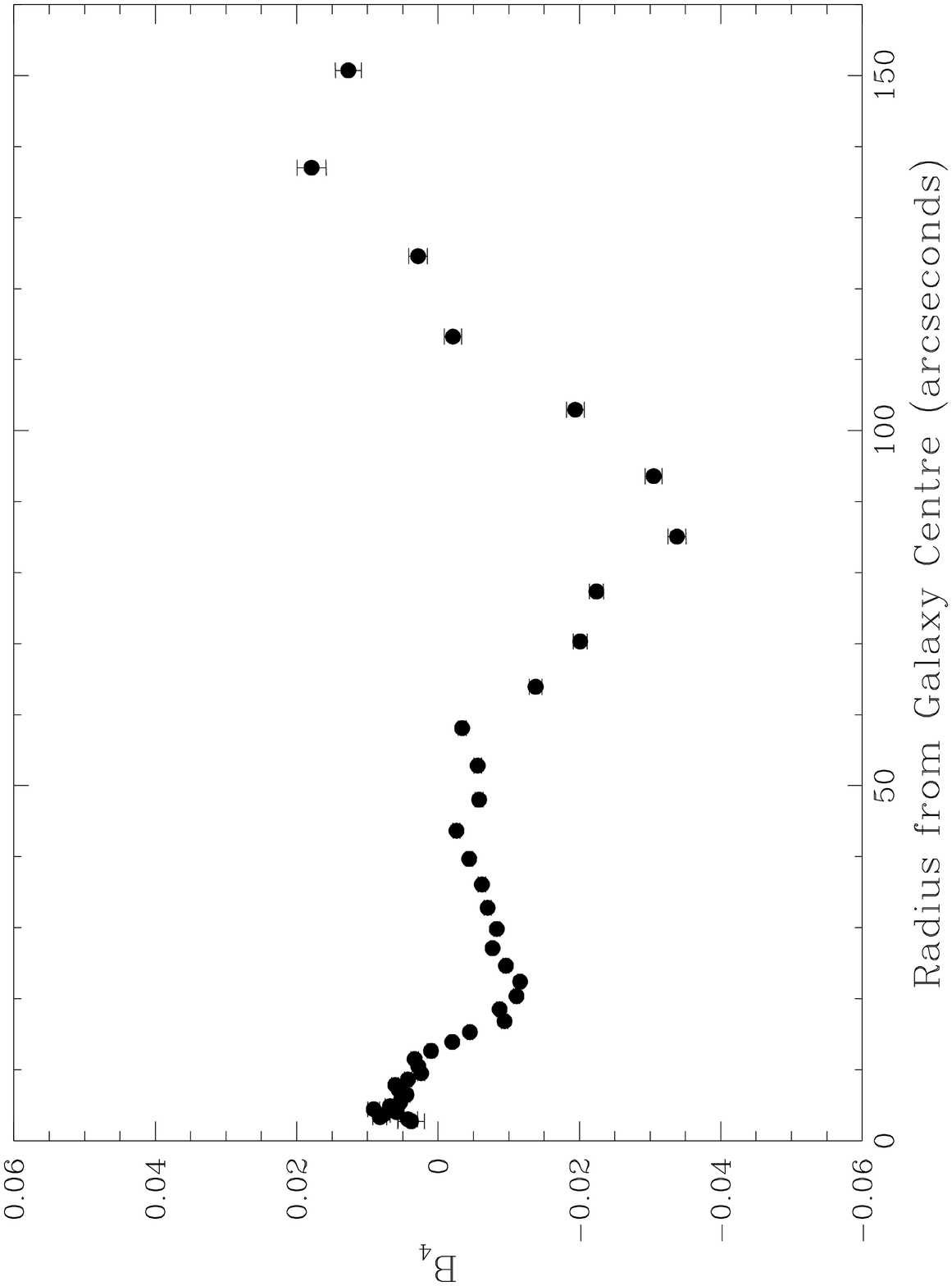},width=8cm,angle=270}
\psfig{{figure=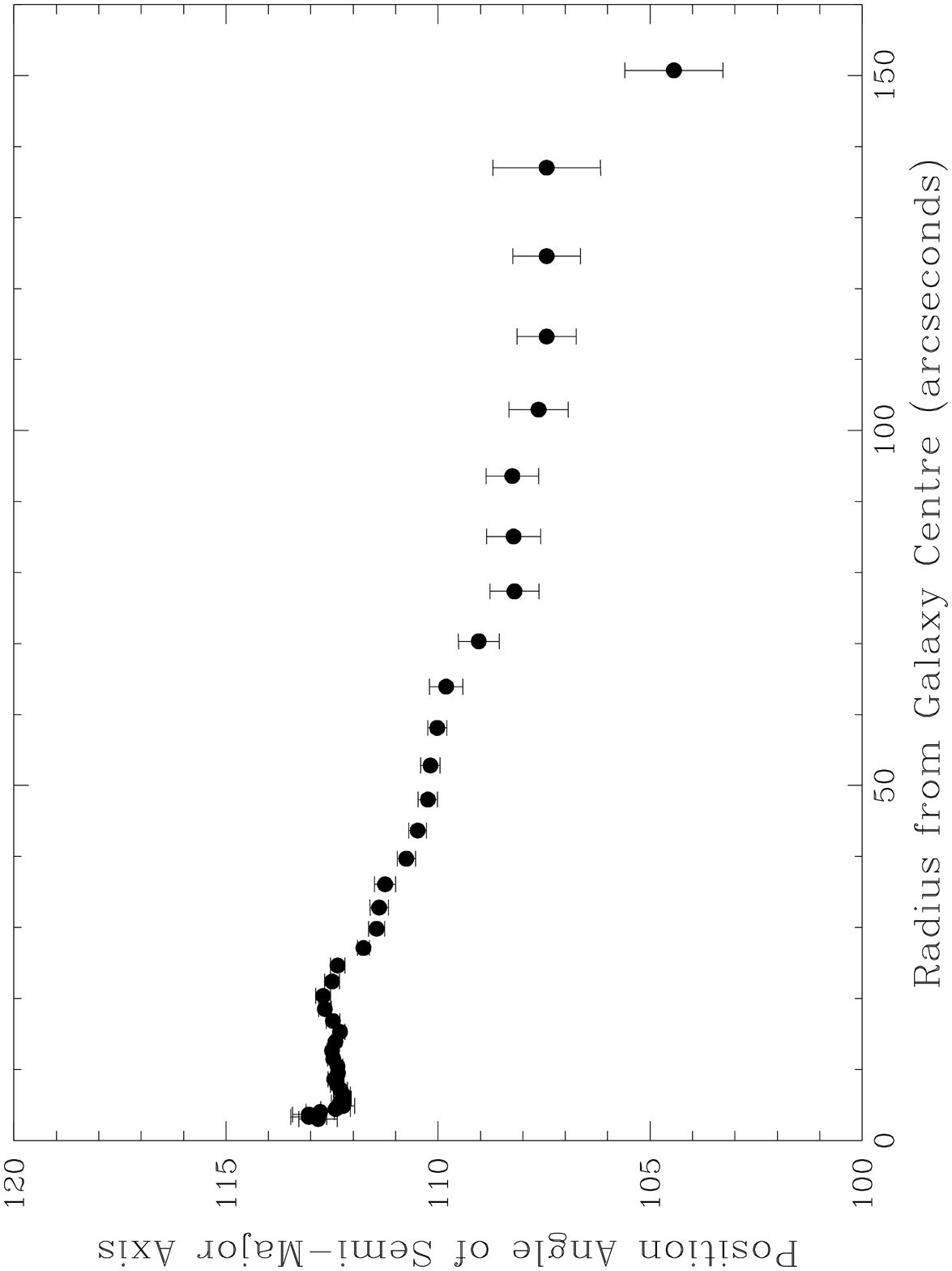},width=8cm,angle=270}
\caption{({\it 9a top}) NGC 7600 cos$B_{4}$ term versus radius and ({\it 9b 
bottom}) 
NGC 7600 position angle of the semi-major axis versus radius.}
\end{figure}

\begin{figure}
\psfig{{figure=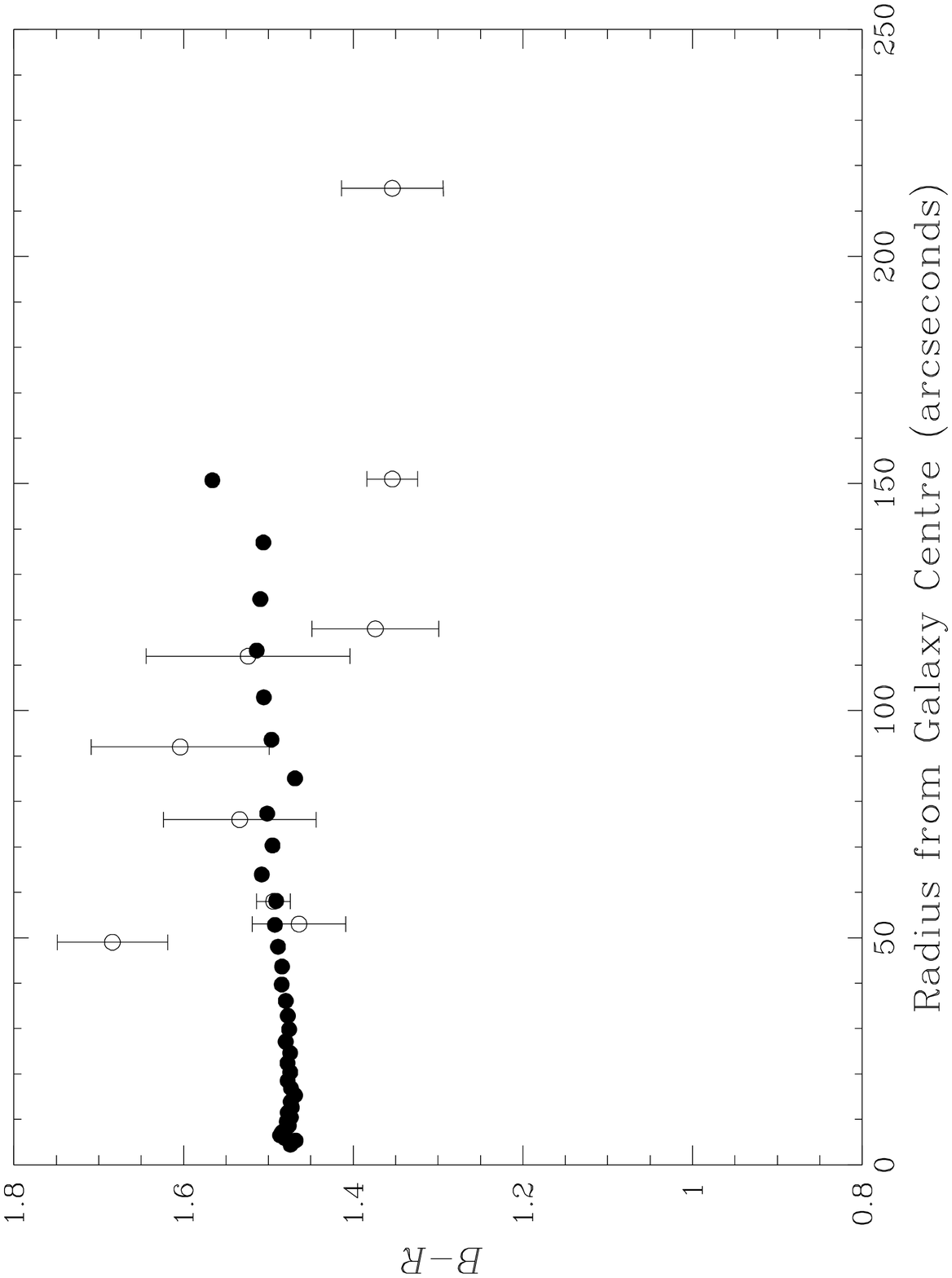},width=8cm,angle=270}
\psfig{{figure=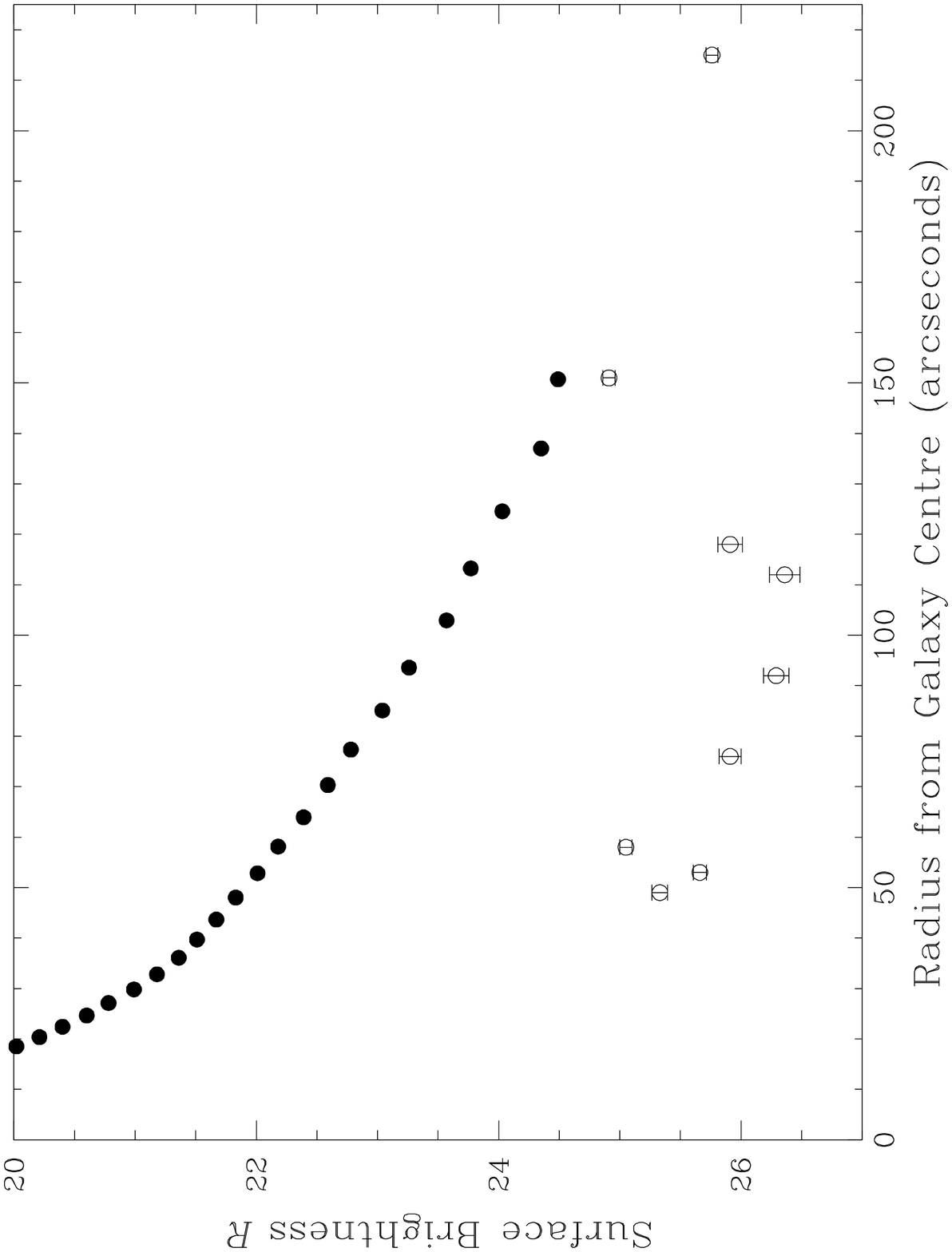},width=8cm,angle=270}
\caption{({\it 10a top}) NGC 7600 $B-R$ of the galaxy and shells 
versus radius and ({\it 10b bottom}) surface brightness of the galaxy 
and shells versus radius. In all cases the galaxy data are represented as 
filled circles, the error being of the order of the symbol size. The shell 
data are open circles with error bars}
\end{figure}

\subsection{$B-R$, Surface Brightness, and Isophotal Parameters}

\subsubsection{NGC 474}

In Figure 5a we plot $R$ surface brightness versus $r^{1/4}$
for NGC 474.  Figures 5b,6ab contain the isophote data obtained for 
NGC 474, namely ellipticity, B$_4$ (the fourth
 cosine term in the Fourier expansion of the azimuthal variation from pure
 ellipse), and position angle 
plotted against radius.
The position 
angle of the semi-major axis and the ellipticity of NGC 474 change quite 
dramatically with radius. The cos B${_4}$ term shows significant scatter and
prevents conclusions from being drawn.

Figure 7a shows the $B-R$ profiles for NGC 474 and its shells.
The galaxy signal falls off to background dominated values 
after 150 arcseconds and these points are not included.  The shell {\it B-R} 
colours seem to follow those of the galaxy; this is more uncertain
for the two outer-most points, since this involves an extrapolation of the
galaxy colours to larger radius. 
It could also be argued that the colours of the {\it inner} shells are 
constant with radius, within the errors; the two outer-most
shells are significantly bluer than the rest.
The $R$ surface brightness of NGC 474 and its shells are plotted in Figure 7b. 
The shell 
surface brightness appears to remain constant with radius, and 
below that of the galaxy within 100 arcsec.  The outer shells, however,
appear to be bluer and it would be useful to measure the galaxy surface
brightness out to this radius. 
The two
outer-most shells also have a higher surface brightness than most of
the inner shells, though not significantly so.  
The fact that the shell 
surface brightness does not follow that of the galaxy argues strongly against
the WIM; we return to this point in Section 5. 

Despite clipping, the question remains as to whether some shell light is 
included in the model images. The residual images would then be an 
underestimate of the true shell light, which if a function of radius, would
effect the final shell colours.
With no galaxy subtraction, the bright shell at a radius of 202 arcseconds 
has a $B-R$ of 1.05 $\pm 0.05$. This value with associated error being 
obtained using the procedure
outlined in section 3.3. From Table 3, we see that this is not a 
significant change within the errors. This shell has a high surface brightness
 and is furthest from the galaxy centre, so we believe that a realistic upper
limit for the systematic uncertainty in the shell colours introduced by the
possible inclusion of shell light in the galaxy model is 0.5 in $B-R$. The 
effect will be smaller for NGC 7600, which has fainter shells. 
Despite some scatter in the points, the isophote plots 
do not seem to show features at the radii of the shells which is a strong
point for proving their exclusion in the model. However, NGC 474 has
a large, high surface-brightness shell at a radius of 55 arcseconds. A model 
run with this shell feature masked out significantly reduced the boxiness 
at that radius. Specifically the B$_4$ parameter was reduced from -0.04 to 
-0.01 with errors of the order $\pm 0.005$. It would seem this shell is 
responsible for the observed boxiness. The position angle and ellipticity were
unaffected.

\subsubsection{NGC 7600}

In Figure 8a, we plot the $R$ surface brightness versus $r^{1/4}$ for NGC 7600.
Figures 8b,9ab plot the isophotal data for NGC 7600. The position 
angle changes little with radius, while the ellipticity rises
sharply within 10 arcsec, and flattens out at $\sim$ 0.55 at larger radius.
The cosB$_4$ term is negative (boxy) at most radii.

Figure 10a shows the $B-R$ for the galaxy and shells. The galaxy $B-R$ is 
flat over the radius plotted (again, the galaxy signal 
is dominated by sky after 150 arcseconds and these points are not included). 
The inner shells have roughly the same colour as the galaxy, while the
three outer-most shells are bluer than the galaxy, with $B-R$ $\simeq$ 1.35. 
Given the errors, it could
be argued that there is a general trend that the shells become
bluer with increasing radius.
The shell and galaxy surface brightness are shown in Figure 10b. The shells in
 NGC 7600 have a surface brightness roughly constant with radius, and do not 
follow the galaxy surface brightness at all (except for possibly the two 
outer-most points). 

\subsection{Associated Fine Structure and Environment}

\subsubsection{NGC 474}

NGC 474 has a peculiar kinematic structure in its core (Balcells, Hau \& 
Carter 1998), with the core rotating about an axis intermediate between photometric 
major and minor axes.  There is also evidence in the rotation curves of this 
galaxy for kinematic shells, of the type also seen in NGC 7626 (Balcells \& 
Carter 1993).  The shell system of NGC 474 was classed as a type II by 
Prieur (1988), meaning that the shells have no preferred orientation 
around the galaxy.  From the residual image in Figure 3 we see that the 
two outer most shells appear to be linked to each other by a `tidal feature'
which crosses the center of the galaxy. 
NGC 474 is a close partner of the smaller spiral NGC 470. Both have 
identical recessional velocities and are at a projected separation of 300 
arcseconds (60 kpc for H$_0$=75 km/sec/Mpc). 
NGC 470 is undergoing an intense nuclear starburst and has a weak bar 
(Freidli et al 1996). It has strong central
CO emission whereas NGC 474 is weaker (Sofue et al 1992). 
The two are linked via an HI tidal bridge, and NGC 470 appears to be in orbit 
about NGC 474 as the HI mimics a `Magellanic stream' around NGC 474 
(Schiminovich et al 1997). 

\subsubsection{NGC 7600}

Dressler \& Sandage (1983) studied the kinematics in the central 20 arcsec of 
NGC 7600.  Although the uncertainties are large, the
bulge $\frac{v}{\sigma}$ versus ellipticity places NGC 7600 close 
towards the theoretical line for a prolate elliptical (see their
Figure 2).  Based on its 
morphology, especially the extended envelope, it has been
classified as an S0.  Dressler \& Sandage claim that the low 
rotation of NGC 7600 implies that it is closer to an elliptical, with its 
support and flattening arising from a large velocity dispersion.  They 
coined the phrase `diskless S0' to describe NGC 7600 and similar galaxies. 
Schweizer \& Seitzer (1988) noted that the ripples of NGC 7600 interleave 
with radius, and concluded that they probably arose from an external 
origin.   We give NGC 7600 the classification of type I, based on this 
interleaving of shells aligned along the major axis.  On inspection 
of the surrounding area, there appears to be no companion to NGC 7600.

\section{Discussion}

Here we discuss our results in light of models of shell formation.

\subsection{NGC 474}

Hernquist \& Quinn (1988) showed that mass transfer, rather than complete 
mergers, were also capable of producing shell structures.  They also noted 
that the efficiency of tidal stripping is greatly reduced in retrograde 
encounters.  Longo \& de Vaucouleurs (1983) quote a mean outer-disk 
colour for NGC 470 of $B-V$ = 0.68.  We transform this to $B-R$ = 1.00, 
 $(\frac{\sigma}{\sqrt{N}})$ $\sim0.08$ 
based on observations of many disk 
galaxies in different filters by de Jong (1995).  If (some of) 
the shells of NGC 474 
are indeed stripped matter from NGC 470, then we must take into account 
subsequent passive stellar evolution (which would redden the shells) and also
 the 
expected blueing of shells as dust is removed during the interaction.  
We will assume for the 
moment that the two processes cancel each other.  (It may also be the case, 
as suggested by the HI maps of the two systems, that NGC 470 
is in orbit about NGC 474 and thus periodically replenishes the outer 
shells; the colours of the (outermost) shells would then also depend on the 
orbital timescale, further complicating the issue). 
Within our uncertainties, the two outer-most shells have colours
consistent with being stripped matter from 
NGC 470 ($B-R$ of 0.91 and 0.99). The inner shells however have a 
mean $B-R$ of 1.27 $\pm$ 0.04, not compatible with the outer-disk colour
of NGC 470.
The discovery of kinematic shells and a peculiar core deep 
within the potential well of NGC 474 is important in this context. It is 
very unlikely that the mass transfer of material could form shells at 
very small radii, and which would exist for the timescales required. 
Kinematically-distinct core (KDC) ellipticals have central regions that 
rotate rapidly, and often in the opposite direction to the stars 
in the outer parts of the galaxy.  KDCs are believed to be result of a 
merger (Illingworth \& Franx 1989), either the accretion 
of a small secondary (Balcells \& Quinn 1990) or a major merger of two nearly 
equal-mass disk galaxies (Hernquist \& Barnes 1991).  Forbes (1992) found 
that in a sample of galaxies with peculiar cores almost all had shells, 
implying that shells and KDCs may form in a similar way, namely via mergers 
(although the role of major mergers in shell formation is poorly understood).

In the WIM (Thomson 1991), the shell system of NGC 474 was formed 
via a weak interaction 
with NGC 470.  Mass transfer to some degree is also compatible with the WIM,
and could explain the origin of the outer shells. 
The similarity of the galaxy and shell $B-R$ profiles in NGC 474 is
in agreement with the WIM, but the fact that the shells' surface brightness
does {\it not} follow that of the galaxy is a very strong (perhaps fatal)
argument {\it against} the WIM.
Hau \& Thomson (1994) showed how a fly-by interaction can form a decoupled 
core by spinning up the envelope of the galaxy. How this would effect the thick
disk is not known but we would expect it to be heated up thus supressing shell
formation. It seems therefore that in order to produce the peculiar core and 
the shells a merger would still be needed even in this scenario.

It seems plausible to use an accretion event to explain the inner shells of
NGC 474 and the peculiar core kinematics. More speculatively, the
two outer-most shells may be the result of mass transfer from NGC 470.

\subsection{NGC 7600}

With no close companions and no evidence for other peculiar structure 
apart from the shells, NGC 7600 is a much cleaner system than
NGC 474.
NGC 7600 is, like NGC 3923, an example of the expected result of a minor 
merger between a prolate elliptical and a dwarf elliptical
on a near radial orbit
(Dupraz \& Combes 1986).  This 
indeed seems to be consistent with the data.  First, the shell system
 is aligned 
along the apparent major axis and appears interleaved.  Second, although
tentative the $B-R$
 of the outer shells 
appear different to the $B-R$ of the galaxy although the inner shells are 
consistent with the $B-R$ of the galaxy. It is difficult to know whether
a gradient exists in the $B-R$ of the shells due to the low surface brightness
of the shells coupled with the missing values of some. On 
balance, however, we believe that the $B-R$ profile and surface brightness of 
the shells point towards them being formed from the stars of a merged 
secondary.

\section{Conclusions and Future Work}

We have presented the first results from our investigation into the 
photometric colours of shell galaxies.  Once reduction and calibration of the 
rest of the sample is complete we hope to make some statements concerning the 
generic properties of these galaxies.  
A key result from the present study of NGC 474 and NGC 7600 is that
the shell surface brightness is roughly constant with radius.  This is
a strong argument against the Weak Interaction Model (Thomson 1991),
which predicts that the surface brightness profile of the shells should
follow that of the galaxy.   Wilkinson et al. (1998/1999) also argue 
against the WIM from their study of the shell system of 0422-476. 

Our hypotheses for the two galaxies is as follows.  NGC 474 may have two 
families of shells. The outer-most 
shells were possibly formed from tidally liberated material from its 
interacting companion NGC 470, while the inner shells formed by a minor merger 
which is now shaping the galaxy's core.  
For NGC 7600, 
our results favour shell formation from a merger/accretion of a smaller 
companion. 

Unfortunately, the various formation models, chiefly mergers and the WIM,
are not developed enough to allow unique, testable predictions for shell
morphologies, surface brightnesses and colours.  For instance, the shell
colours of NGC 474 and NGC 7600 do not allow us to distinguish between a
WIM and a merger model.  Detailed numerical simulations of shell formation
in the various scenarios, incorporating dynamical friction and 
realistic treatments of gas and dust,
are needed to differentiate between the different models, and to learn more
about the progenitors.  On the observational side, shell {\it kinematics}
holds considerable promise.  For instance, Merrifield \& Kuijken (1998)
showed that the measurement of shell kinematics in regular, aligned systems
provides a means for determining the gravitational potential of their
host galaxies out to large radii.   Such measurements will become feasible
with integral-field spectrographs on 8--10m class telescopes.

\section{Acknowledgements}

AJT would like to thank Bob Thomson for initiating the project and for 
many helpful ideas and interesting discussions. We would like to thank 
Jim Collett for contributions to the discussion.

This research has made use of the NASA/IPAC Extragalactic Database (NED)
which is operated by the Jet Propulsion Laboratory, California Institute of 
Technology, under contract with the National Aeronautics and Space Administration.     

The Isaac Newton Telescope is operated on the island of La Palma by the 
Royal Greenwich Observatory in the Spanish Observatorio del Roque de los 
Muchachos of the Instituto de Astrofisica de Canarias.

\section{References}

\begin{trivlist}
\item[] Arp, H. 1966, in {\it Atlas of Peculiar Galaxies}, Pasadena, 
California Institute of Technology
\item[] Balcells, M. 1997, ApJ, 486, L87
\item[] Balcells, M., \& Carter, D. 1993, Astron. Astrophys, 279, 376
\item[] Bender, R., Dobereiner, S \& Mollenhoff, C., 1988. Astr. Astrophys. 
Suppl, 74, 385
\item[] Bender, R., 1990, in {\it Dynamics and 
Interactions of Galaxies,} p.232, ed. R. Wielen (Springer, Berlin)
\item[] Binney, J., \& Petrou, M. 1985, MNRAS, 214 449
\item[] Carter, D., preprint for IAU 186, 1997
\item[] Carter, D., Allen, D. A., \& Malin, D. F. 1982, Nature, 295, 126
\item[] Carter, D., Thomson, R. C., \& Hau, G. K. T. 1998, MNRAS, 294, 
182
\item[] de Jong, R. S. 1995, {\it Thesis}, Groningen, Netherlands.
\item[] Dressler, A., \& Sandage, A. 1983, ApJ, 265, 664
\item[] Dupraz, C., \& Combes, F. 1986, A\&A, 166, 53 
\item[]	Dupraz, C., \& Combes, F. 1987, A\&A, 185, L1
\item[] Fort, B. P., Prieur, J-L., Carter, D., Meatheringham, S. J., \& 
Vigroux, L. 1986, ApJ, 306, 110
\item[] Forbes, D. A., Reitzel, D. B., Williger, G. M. 1995, 
Astron. J, 109(4), 1576
\item[] Forbes, D. A., \& Thomson, R. C. 1992, MNRAS, 254, 723
\item[] Friedli, D., Wozniak, H., Rieke, M,. Martinet, L., \& Bratschi, P.
 1996, A\&Asupp, 118, 461
\item[] Heisler, J., \& White, S. D. M. 1990, MNRAS, 243, 199 
\item[] Hibbard, J. E., \& Mihos, J. C. 1995, Astron. J, 110, 140
\item[] Hernquist, L., \& Quinn, P. J. 1988, ApJ, 331, 682
\item[] Hernquist, L., \& Quinn, P. J. 1989, ApJ, bf 342, 1
\item[] Hernquist, L., \& Spergel, D. N. 1992, ApJ, 399, L117
\item[] Jedrzejewski, R. I. 1987, MNRAS, 226, 747
\item[] Jorgensen, I. 1994, PASP, 106, 967
\item[]	Landolt, A. U., 1992, AJ, 104(1), 340
\item[] Longo, G \& de Vaucouleurs, A. 1983, Univ. Tex. Monogr. Astron. No.3
\item[]	Malin, D. F., \& Carter, D. 1983, ApJ, 274, 534
\item[] Malin, D. F., \& Carter, D. 1980, Nature, 285, 643
\item[] Merrifield, M., \& Kuijken, K. 1998, MNRAS, 29, 1292 
\item[] Nieto, J-L., Bender, R., Arnaud, J., \& Surma, P. 1991, Astr. 
Astrophys, 244, L25
\item[] Penereiro, J. C., de Carvalho, R. R., Djorgovski, S., \& Thompson, D.
  1994, A\&A Supp, 108, 461
\item[] Prieur, J-L., 1988, {\it Thesis}, Toulouse, (France)
\item[] Prieur, J-L., 1990, in {\it Dynamics and Interactions of 
Galaxies,} p.72, ed. R. Wielen (Springer, Berlin)
\item[]	Quinn, P.J. 1984, ApJ, 279, 596
\item[]	Schiminovich, D., Van Gorkom, J. H \& Van der Hulst, J. M. 1997, IAU 
S186
\item[]	Schombert, J. M., \& Wallin, J. F. 1987, AJ, 94(2), 300
\item[] Schweizer, F. 1980, ApJ, 237, 303
\item[]	Schweizer, F., \& Seitzer, P. 1988, ApJ, 328, 88
\item[]	Seitzer, P., \& Schweizer, F. 1990, in {\it Dynamics and 
Interactions of Galaxies,} p.270, ed. R. Wielen (Springer, Berlin)
\item[]	Thomson, R. C., \& Wright, A. E. 1990, MNRAS, 247, 122
\item[]	Thomson, R.C. 1991, MNRAS, 257, 689
\item[] Visvanathan, N., \& Sandage, A. 1977, ApJ, 216, 214
\item[] Wilkinson, A., Prieur, J.-L., Lemoine, R., Carter, D., Malin, D., 
Sparks, W. B. 1998, preprint
\end{trivlist}

\end{document}